# Giant Anisotropic Magnetoresistance and Planar Hall Effect in the Dirac Semimetal Cd$_3$As$_2$

Hui Li[1], Huan-Wen Wang[2], Hongtao He[3], Jiannong Wang[1]*, Shun-Qing Shen[2]*

[1]*Department of Physics, the Hong Kong University of Science and Technology, Clear Water Bay, Hong Kong, China*

[2]*Department of Physics, the University of Hong Kong, Pokfulam Road, Hong Kong, China*

[3]*Department of Physics, Southern University of Science and Technology, Shenzhen, Guangdong 518055, China*

**Anisotropic magnetoresistance is the change tendency of resistance of a material on the mutual orientation of the electric current and the external magnetic field. Here, we report experimental observations in the Dirac semimetal Cd$_3$As$_2$ of giant anisotropic magnetoresistance and its transverse version, called the planar Hall effect. The relative anisotropic magnetoresistance is negative and up to -68% at 2 K and 10 T. The high anisotropy and the minus sign in this isotropic and nonmagnetic material are attributed to a field-dependent current along the magnetic field, which may be induced by the Berry curvature of the band structure. This observation not only reveals unusual physical phenomena in Weyl and Dirac semimetals, but also finds additional transport signatures of Weyl and Dirac fermions other than negative magnetoresistance.**

Anisotropic magnetoresistance (AMR) was first discovered by Thomson in 1857 and was observed in many ferromagnetic metals [1]. It is closely associated with changes of magnetization relative to the current. Its transverse version shows a transverse current or voltage in response to the longitudinal current flow and an applied in-plane magnetic field, called the planar Hall effect (PHE) [2,3]. Usually the relative AMR is weak, less than 1% or at most up to a few percent in some ferromagnetic metals and half-metallic ferromagnets [4]. Recently, negative longitudinal magnetoresistance and non-saturated linear out-of-plane perpendicular magnetoresistance and in-plane transverse magnetoresistance were reported in a series of newly discovered topological semimetals [5-16]. While negative magnetoresistance is possibly associated with the chiral anomaly of the Weyl fermions in an electric field and a magnetic field [17-20], linear out-of-plane perpendicular magnetoresistance and in-plane transverse magnetoresistance illustrate high anisotropy of magneto-transport in topological semimetals. This is a rare property for a paramagnetic metal. The AMR and PHE have started to attract a lot of theoretical studies in topological semimetals [21,22] and other topological materials [23-25].

We denote transverse resistivity by $\rho_\perp(B)$ when the magnetic field **B** is perpendicular to the electric current density **j**, i.e., $\mathbf{j} \cdot \mathbf{B} = 0$, and longitudinal resistivity by $\rho_\parallel(B)$ when the magnetic field is parallel to the electric current density, i.e., $\mathbf{B} \parallel \mathbf{j}$. Usually transverse resistivity is larger than longitudinal resistivity, i.e., $\rho_\perp(B) \geq \rho_\parallel(B)$, in Weyl and Dirac semimetals. The equality holds only for $B = 0$. Thus, the resistivity is very sensitive to the angle between the electric current density and the magnetic field. In general, the AMR and PHE can be well described by a formula between the electric field **E** and the electric current density in a vector form as below,

$$\mathbf{E} = \rho_\perp \mathbf{j} + (\rho_\parallel - \rho_\perp)\frac{\mathbf{B}\mathbf{B}\cdot\mathbf{j}}{B^2} + \rho_\perp \chi \mathbf{B} \times \mathbf{j}, \qquad (1)$$

where $\chi$ is the mobility of the charge carriers. Assuming **j** and **B** construct an $x - y$ plane [see Fig. 1(a)], the in-plane field-dependent resistivity $\rho_{ij} = \rho_\perp \delta_{ij} + (\rho_\parallel - \rho_\perp)B_i B_j/B^2$ with $i,j = x,y$. In-plane diagonal or longitudinal resistivity is highly anisotropic as a function of the angle $\varphi$ between the magnetic field and electric current density, $\rho_{xx} = \frac{\rho_\parallel + \rho_\perp}{2} + \frac{\rho_\parallel - \rho_\perp}{2}\cos 2\varphi$. In-plane off-diagonal resistivity leads to a non-zero electric field that is normal to the electric current density but parallel to the magnetic field. In-plane off-diagonal resistivity is called planar Hall resistivity, $\rho_{xy} = \frac{\rho_\parallel - \rho_\perp}{2} \sin 2\varphi$. It is worth emphasizing that $\rho_{xx}(\varphi)$ and $\rho_{xy}(\varphi)$ have identical forms to that for ferromagnetic metals [1].

In our previous magnetotransport study of Dirac semimetal Cd$_3$As$_2$ microribbons [8], we have successfully observed the carrier density dependence of non-saturating positive magnetoresistance in out-of-plane perpendicular magnetic fields and negative longitudinal magnetoresistance in parallel magnetic fields. Here, we present further in-plane magnetotransport results on the AMR and PHE in Dirac semimetal Cd$_3$As$_2$ microribbons. Our experimental results are in excellent agreement with the relation between the magnetic field and electric current density in Eq. (1), which can be derived from the field-dependent current induced by the chiral anomaly of Weyl and Dirac fermions, or the Berry curvature in conventional and topological metals in the semiclassical theory. This implies that the observed AMR and PHE in our Cd$_3$As$_2$ microribbons is associated with the physics of Berry curvature intrinsic to the Dirac semimetal Cd$_3$As$_2$.

Detailed growth and structural charaterizations of Cd$_3$As$_2$ microribbons can be found in our earlier work [8]. In brief, a Cd$_3$As$_2$ microribbon was grown by the chemical vapor deposition method on Si (001) substrates and Ar gas was used as a carrier gas. The furnace was gradually heated up to 750 °C in 20 min and the Ar flow was kept as 100 sccm (sccm denotes cubic centimeter per minute at standard temperature and pressure) during the growth process. The duration of the growth is 60 min, and then it cools down to room temperature naturally. To study the magnetotransport properties, a Hall bar structure was fabricated with standard e-beam lithography and lift-off processes. Al/Au/Cr electrodes with thicknesses of 500 nm / 175 nm / 25 nm were deposited using thermal evaporation and e-beam evaporation methods. The transport properties of the devices were then measured in a Quantum Design physical properties measurement system (PPMS) with the highest magnetic field up to 14 T.

Figure 1(a) shows the optical image of a typical Cd$_3$As$_2$ device studied in this work. The width $w$ is about 5 μm, and the inter-voltage-probe distance for $V_{xx}$ and $V_{xy}$ is about 10 and 2 μm, respectively. The ribbon thickness $t$ is about 592 nm according to the atomic force microscopy measurement shown in Fig. 1(b) and the measured height profile in Fig.

1(c). The current **I** is applied along the longitudinal direction (*x*-direction) of the $Cd_3As_2$ microribbon, as indicated in Fig. 1(a). The carrier concentration and mobility is in the order of $10^{17}/cm^3$ and $10^4$ cm$^2$/Vs, respectively, for temperatures below 50 K. The Fermi energy $E_F$, defined as the energy difference between the Fermi level and the Dirac point, is estimated to be about 88 meV above the Dirac point based on the known Fermi velocity $v_F \sim 10^6$ m/s for $Cd_3As_2$ [26] (see Sec. A in the Supplemental Material [27] for details). Figure 1(d) shows the in-plane longitudinal magentoresistance ($MR_{xx} = \frac{R_{xx}(B)-R_{xx}(0)}{R_{xx}(0)} \times 100\%$) measured at $T = 2$ K with various angles $\varphi$ between the applied magnetic field and current directions in the $x-y$ plane (see the inset). When the magnetic field (**B**-field) is parallel to the applied current (**I**), i.e. $\varphi = 0°$, a pronounced negative $MR_{xx}$ is observed in low magnetic fields. When $\varphi$ increases, negative $MR_{xx}$ vanishes and an evident positive $MR_{xx}$ is observed. It reaches the maximum value of ~ 275% at 14 T when the **B**-field is transverse to **I** ($\varphi = 90°$). Figure 1(e) shows the $MR_{xx}$ curves measured at $\varphi = 0°$ at the indicated temperatures. Negative $MR_{xx}$ has been observed in a wide temperature range, and the largest negative $MR_{xx}$ of ~ -21% is observed at $T = 50$ K and $B = 7$ T. However, negative $MR_{xx}$ is not observable when the temperature further increases to above 200 K. Similar $MR_{xx}$ behaviors have been observed in other $Cd_3As_2$ microribbon devices. Such negative $MR_{xx}$ has been studied in detail and attributed to the chiral anomaly in our previous work [8]. Similar observations of negative $MR_{xx}$ with **B** ∥ **I** have also been reported and are considered as a signature of the chiral anomaly of topological semimetals [5-7,9-12].

According to Eq. (1), the longitudinal resistance (diagonal) $R_{xx}$ and the planar Hall resistance (off-diagonal) $R_{xy}$, defined as $R_{xy} = \frac{V_{xy}}{I}$, change systematically as a function of the rotating angle $\varphi$. Figure 2(a) is a schematic illustration of the PHE in the $Cd_3As_2$ microribbon devices, the current **I** is applied along the longitudinal direction of the $Cd_3As_2$ microribbon, and the **B**-field is rotated in the $x-y$ plane. In the experiment, a misalignment between the actual rotation plane and the $Cd_3As_2$ microribbon plane may exist. This could result in a finite ordinary Hall resistivity component in the measured planar Hall resistivity $R_{xy}$. Fortunately, the ordinary Hall component is antisymmetric to the **B**-field directions and can be readily eliminated by summing the measured $R_{xy}$ in both positive and negative **B**-field directions. Figure 2(b) shows the symmetrized **B**-field dependence of the $R_{xy}$ measured at different rotating angles $\varphi$ at 2 K. When $\varphi$ varies from 0° to 90°, the magnitude of $R_{xy}$ increases first and then decreases as expected in the PHE discussions of Eq. (1).

Figure 2(c) shows the symmetrized angular-dependent $R_{xy}$ and $R_{xx}$ measured at 2 K and 5 T. Both the measured $R_{xy}$ and $R_{xx}$ show a 180° periodic angular dependence, which is not expected for a non-magnetic and isotropic solid but is in agreement with Eq. (1). We fit the $R_{xy}$ using the equation $R_{xy} = \gamma \frac{R_\parallel - R_\perp}{2} \sin 2\varphi$ derived from Eq. (1), where $R_\parallel$ and $R_\perp$ is the longitudinal resistance when $\varphi$ equals to 0 and 90°, respectively, and $\gamma$ is the geometric ratio of the width to the length of the Hall bar device. Remarkably, as indicated by the red line in the top panel in Fig. 2(c), the measured $R_{xy}$ can be well fitted by the equation, demonstrating the existence of the PHE in this nonmagnetic material. Moreover, the measured $R_{xx}$ can also be well fitted by the equation $R_{xx}(B,\varphi) = \frac{R_\parallel + R_\perp}{2} + \frac{R_\parallel - R_\perp}{2} \cos 2\varphi$, as indicated by the red line in the bottom panel in Fig. 2(c). Such a periodic resistance oscillation is a peculiar characteristic of the AMR, as will be discussed below. The same PHE and AMR features have been observed in other Cd$_3$As$_2$ microribbon devices, as can be seen in Sec. B in the Supplemental Material [27].

Prior to discussing the AMR effect, we revisit the $R_{xx}$ − **B** curves at different rotating angle $\varphi$, as shown in Fig. 3(a). According to Eq. (1), the longitudinal resistance $R_{xx}$ follows $R_{xx}(B,\varphi)$. Theoretically, we can calculate the $R_{xx}$ value at any arbitrary angle $\varphi$ if the $R_{xx}$ values at $\varphi = 0°$ ($R_\parallel$) and 90° ($R_\perp$) are known. As shown in Fig. 3(a), the measured $R_{xx}$ − **B** curves at $\varphi = 0°$, 30°, 60° and 90° in a magnetic field range of ±5 T are plotted. The red lines are the calculated results of $R_{xx}(B,\varphi)$ using the measured curves at $\varphi = 0°$ and 90°. The yielded $\varphi$ is 27° and 55°, respectively, which is very close to the experimental set value of 30° and 60°. The difference may be caused by the misalignment between the actual rotation plane and the Cd$_3$As$_2$ microribbon plane. Figure 3(b) shows the symmetrized angular-dependent $R_{xx}$ at the **B**-fields indicated and 2 K. The AMR effect can be seen clearly at different **B**-fields and can be well described by $R_{xx}(B,\varphi)$ [red lines in Fig. 3(b)].

More remarkably, the observed AMR here shows anomalously larger $R_\perp$ than $R_\parallel$. Thus, the AMR ratio, defined as $\frac{R_\parallel - R_\perp}{R_\perp} \times 100\%$[4], is negative for the Cd$_3$As$_2$ microribbon device at 2 K, as shown in Fig. 3(c). With increasing **B**-fields, the magnitude of the AMR ratio increases monotonically and saturates at ~ 68% around $B$ = 10 T. In Fig. 3(c), indicated as solid symbols, the saturated AMR ratios for ferromagnetic metals CoMnAl, NiFe, and Fe$_4$N and half-metallic ferromagnet La$_{0.7}$Sr$_{0.3}$MnO$_3$ are replotted from Ref. [4] for comparison. As can be seen, the saturated AMR ratio for the Cd$_3$As$_2$ microribbon device is one or two orders of magnitude larger than that for ordinary ferromagnetic metals and half-metallic ferromagnets. This giant and negative AMR is a striking feature in topological Weyl and Dirac semimetals. Moreover, the AMR amplitude follows a quadratic **B**-field dependence at a small **B**-field regime ($B$ < 1.0 T) as theoretically expected (see Sec. C in the Supplemental Material [27] for details). However, the AMR amplitude deviates from the quadratic function at a higher **B**-field above 3.5 T, which may be caused for multiple reasons and is still under theoretical investigation.

Figure 3(d) shows the angular-dependent magnetoconductance. The longitudinal magnetoconductance is not simply proportional to $\cos^2 \varphi$. The non-sinusoidal feature becomes more evident when increasing the **B**-field from 1 to 10 T, demonstrating that the measured AMR increased with the magnetic field. This effect can be well understood from the AMR and PHE [21]. In this case, the conductance $G$ is given as $R^{-1}$, and the relative longitudinal magnetoconductance is then given by

$$\frac{G - G_\perp}{G_\perp} = \frac{\cos^2 \varphi}{\frac{R_\parallel}{R_\perp - R_\parallel} + \sin^2 \varphi}. \qquad (2)$$

The correction of $\sin^2 \varphi$ in the denominator reflects the in-plane transverse voltage induced by the applied **B**-field or the PHE. This is another peculiar feature of the AMR. We plug the experimental value of $R_\parallel$ and $R_\perp$ into Eq. (2) to reproduce the angular-dependent magnetoconductance [red lines in Fig. 3(d)], which shows very good agreement with the experimentally measured ones [open circles in Fig. 3(d)]. Thus the PHE is attributed to this angle narrowing effect. This angle narrowing effect was also observed in a previous experiment [6], which implies the existence of PHE in Na$_3$Bi.

The AMR and PHE can be attributed to a **B** field-dependent current given by

$$\mathbf{j}_B = \alpha(\mathbf{B} \cdot \mathbf{E})\mathbf{B}. \tag{3}$$

The key feature of Eq. (3) is that the current density is parallel to the magnetic field instead of the electric field. Several mechanisms may produce this type of current. (1). The chiral magnetic effect of Weyl fermions gives a non-zero current density which is proportional to the magnetic field **B**, $\mathbf{j}_{CME} = \frac{e^2}{4\pi^2\hbar^2}\mu_5\mathbf{B}$, where $\mu_5$ is the chemical potential difference between the two Weyl nodes according to the quantum field theory [28,29]. However, when the Weyl fermions are subject to both an electric field **E** and a magnetic field **B**, the chiral anomaly equations for Weyl fermions induce a non-zero value $\mu_5 \approx \frac{e^2\hbar v_F^3 \mathbf{E}\cdot\mathbf{B}\tau_v}{\mu^2}$, where $\tau_v$ is the relaxation time, $v_F$ is the Fermi velocity, and $\mu$ is the chemical potential. Substituting $\mu_5$ into the current equation of the chiral magnetic effect, one obtains $\alpha = \frac{e^2}{4\pi^2\hbar}\frac{e^2 v^3 \tau_v}{\mu^2}$. $\alpha$ is anticipated to be a constant in weak magnetic fields [18] and become field-dependent at strong fields [30]. (2). In the second-order semiclassical theory [31], the Berry curvature in a conventional metal without chiral anomaly can also produce a current along the direction of the **B**-field. It is attributed to the Fermi surface property that highly depends on the geometric quantities such as the orbital magnetic moment. For nonmagnetic metals in the semiclassical regime, the leading-order magnetoresistivity is quadratic of $B$ due to the constraint of time-reversal symmetry and the Onsager's relation [32]. (3). Other possible mechanisms are also proposed in conventional and topological conductors. For example, the electric and magnetic field can produce a helicity imbalance leading to the field-dependent current in a Dirac-like material [33].

In the presence of a magnetic field, the Lorentz force, which deflects the motion of charged particles in a magnetic field, is also one of the main sources to produce magnetotransport in a solid. Considering the drift velocity of charge carriers in a magnetic field and the field-dependent correction to the charge current, the charge current density **j** can be expressed as

$$\mathbf{j} - \chi \mathbf{j} \times \mathbf{B} = \sigma_D \mathbf{E} + \alpha(\mathbf{E}\cdot\mathbf{B})\mathbf{B} \tag{4}$$

where $\sigma_D$ is the isotropic conductivity, and $\chi$ is the mobility. The resulting resistivity is a tensor instead of a scalar after $\mathbf{j}_B$ is included. When the magnetic field is transverse to the electric current density, i.e. $\mathbf{j}\cdot\mathbf{B} = 0$, the transverse resistivity is $\rho_\perp = \frac{1}{\sigma_D}$. When the magnetic field is parallel to the electric current density, i.e. **B** ∥ **j**, the longitudinal resistivity is $\rho_\parallel(B) = \frac{1}{\sigma_D + \alpha B^2}$. The parameter $\alpha$ can then be expressed in terms of $\rho_\parallel$ and $\rho_\perp$ as $\alpha = (\rho_\parallel^{-1} - \rho_\perp^{-1})/B^2$. In practice, $\rho_\parallel$ and $\rho_\perp$ are two physical quantities to be measured experimentally. As a result, $\alpha B^2$ may give rise to a negative magnetoresistivity defined as $\delta\rho_\parallel = \rho_\parallel(B) - \rho_\parallel(0)$. In addition, Eq. (1) can be explicitly derived from Eq. (4) by the vector calculation. All the parameters in Eq. (1) are measurable experimentally.

The excellent agreement between the measured AMR and PHE and that described by Eqs. (1) and (4) reveals the existence of the field-dependent current [given by Eq. (3)] in our $Cd_3As_2$ devices. For the Dirac semimetal $Cd_3As_2$, the conduction and valence bands are inverted near the $\Gamma$ point to form Dirac points and the Lifshitz point, but a linear dispersion persists even when the Fermi energy $E_F$ moves up to 250 meV above the Dirac point [26]. Since the Lifshitz energy is relatively small [34], it is believed that a strong coupling between the conduction and valence bands produces the Berry curvature, which can induce the field-dependent current according to the semiclassical theory [31,33]. If the two bands just touch at one point [35], the chiral anomaly could provide a reasonable mechanism to produce the AMR and PHE [22].

Large in-plane transverse magnetoresistance has not been well understood up to now. In this case, the effect of chiral anomaly is ruled out as the current is normal to the magnetic field. Although Abrisokov found a linear magnetoresistance at a screened Coulomb potential in the quantum limit [36], a linear magnetoresistance was observed even at relatively weak fields in many Weyl and Dirac semimetals, especially those with high mobility [37]. Therefore, the true physical mechanism is still unclear [38]. In this work, a quadratic in-plane transverse magnetoresistance is only measured in very weak fields B < 1 T as shown in Fig. 1(d). Large positive in-plane transverse magnetoresistance clearly deviates from the quadratic behaviors when $B > 2$ T. This cannot be simply explained in the framework of the Drude theory assuming that the relaxation time is independent of field. However, the observed large magnetoresistance indicates that the relaxation time has a large correction in a finite magnetic field. Negative longitudinal magnetoresistance has been discussed in the previous paper [8]. It is worth noting that negative magnetoresistance can be also induced by some mechanisms other than chiral anomaly [31,33,39-43].

Whether these mechanisms can also induce the AMR and PHE is still an open question and deserves further study.

To conclude, we have observed giant and negative anisotropic magnetoresistance and the planar Hall effect in nonmagnetic $Cd_3As_2$ microribbons. Our experimental results are in excellent agreement with the theoretical descriptions and formulas of AMR and PHE. The puzzle of the angle narrowing effect which was first observed in $Na_3Bi$ is also resolved according to the theory of AMR and PHE. Therefore, our work not only reveals unusual physical phenomena in other semimetals, but also find additional transport signatures of other fermions other than negative magnetoresistance. The observed giant AMR and PHE in topological semimetals might also have potential applications in magnetic sensors.

Recently, we became aware of a work on the measurement of the planar Hall effect by Wu et al. [44].

Acknowledgements This work was supported in part by the Research Grants Council of the Hong Kong SAR under Grants No. 16305215, No. 17301116, No. C6026-16W and No. AoE/P-04/08, and in part by the National Natural Science Foundation of China (No. 11574129) and the Natural Science Foundation of Guangdong Province (No. 2015A030313840). The electron-beam lithography facility is supported by the Raith-HKUST Nanotechnology Laboratory at MCPF.

H.L. and H.-W.W. contributed equally to this paper.

* Correspondence and requests for materials should be addressed to S.S. (email: sshen@hku.hk) and J.W. (email: phjwang@ust.hk).

**Figure Captions**

**Figure 1. Cd$_3$As$_2$ microribbon devices and magneto-transport characteristics.** (a) The optical image of the Cd$_3$As$_2$ microribbon device. The current probes **I** and voltage probes $V_{xx}$ and $V_{xy}$ are labeled in orange, green and purple colors, respectively. The current is applied along the $x$ direction as indicated, and the microribbon is rotated in the $x - y$ plane. (b) AFM image and (c) the height profile of the Cd$_3$As$_2$ microribbon in (a). (d) The in-plane longitudinal magnetoresistance (MR$_{xx}$) measured at 2 K with the applied magnetic field (**B**) direction changing from parallel ($\varphi = 0°$) to transverse ($\varphi = 90°$) to the applied current (**I**) direction in the $x - y$ plane. (e) In-plane longitudinal magnetoresistance (MR$_{xx}$) measured at the different temperatures indicated in parallel fields in the $x - y$ plane.

**Figure 2. Planar Hall effect (PHE) in the Cd$_3$As$_2$ microribbon.** (a) Schematic of the PHE in the Cd$_3$As$_2$ microribbon devices. (b) The magnetic field dependence of the $R_{xy}$ measured at 2 K with the applied **B**-field direction changing from parallel ($\varphi = 0°$) to transverse ($\varphi = 90°$) to the applied current direction in the $x - y$ plane. (c) The symmetrized angular dependence of the $R_{xy}$ (top panel) and $R_{xx}$ (bottom panel) measured at 2 K and 5 T. The red lines are the fitting curves using the inset equations, where $\varphi$ is the angle between the **I** and **B**-field in the $x - y$ plane; $R_\parallel$ and $R_\perp$ are the resistance when $\varphi$ is equal to 0 and 90°, respectively; $\gamma$ is the ratio of the width to the length of the Cd$_3$As$_2$ microribbon device.

**Figure 3. Anisotropic magnetoresistance (AMR) in the Cd$_3$As$_2$ microribbon.** (a) The symmetrized longitudinal resistance ($R_{xx}$) measured at 2 K at $\varphi$ indicated, where $\varphi$ is the angle between the **I** and **B**-field in the $x - y$ plane. (b) The angular dependence of the $R_{xx}$ at 2 K and **B**-fields indicated. The red lines in (a) and (b) are fitting curves using $R_{xx} = \frac{R_\parallel + R_\perp}{2} + \frac{R_\parallel - R_\perp}{2} \cos 2\varphi$. (c) The AMR ratio of the Cd$_3$As$_2$ microribbon devices at 2 K as a function of the **B**-fields. The black line is a guide to the eyes. The AMR ratio of ferromagnetic metals, CoMnAl, NiFe, Fe$_4$N, and half-metallic ferromagnet La$_{0.7}$Sr$_{0.3}$MnO$_3$ from Ref. [23] is also indicated for comparison. (d) The angular dependence of the in-plane

longitudinal conductance $G_{xx}$ at 2 K and **B**-fields indicated. The red lines are fitting curves using Eq. (3).

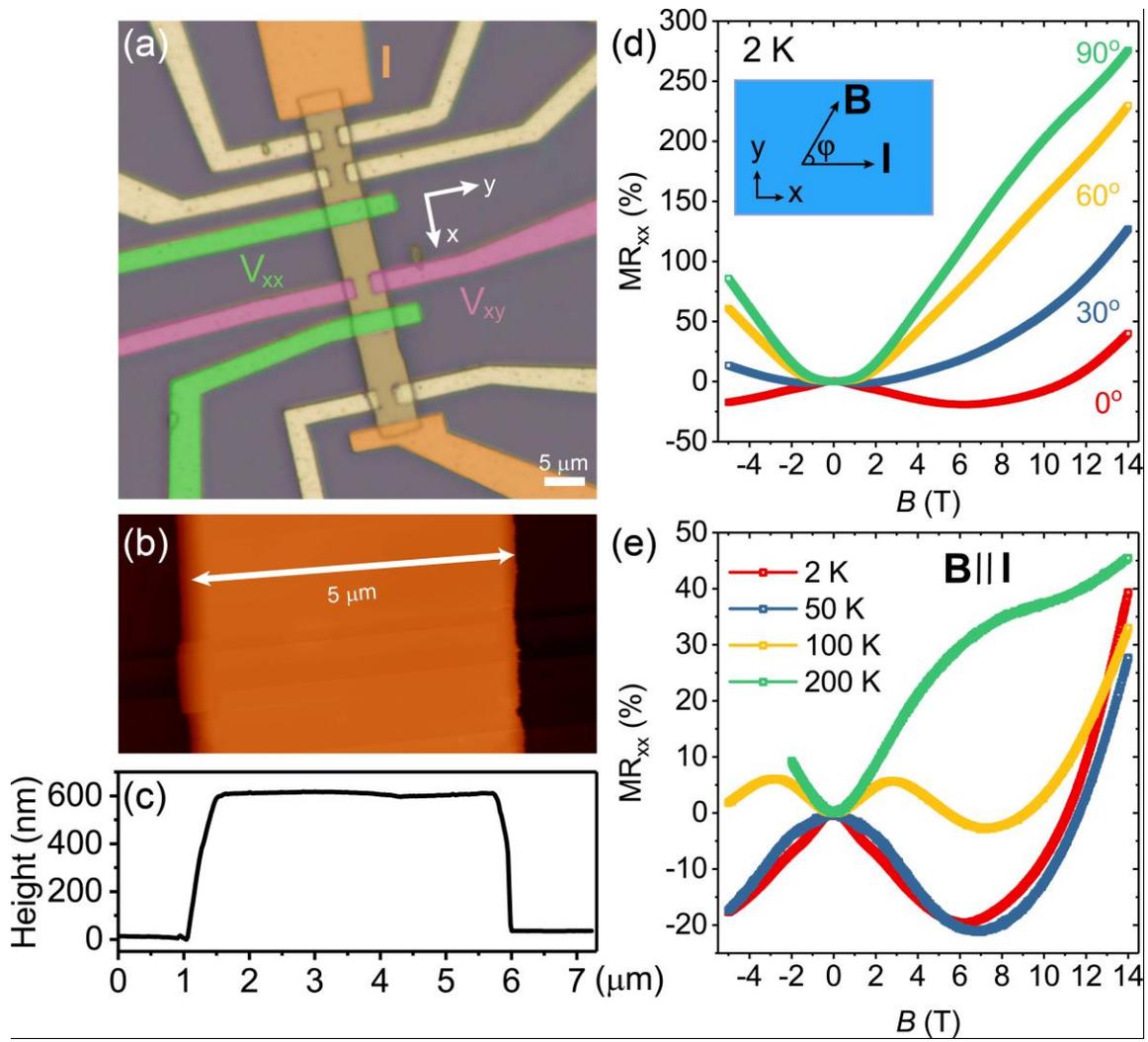

**Figure 1**

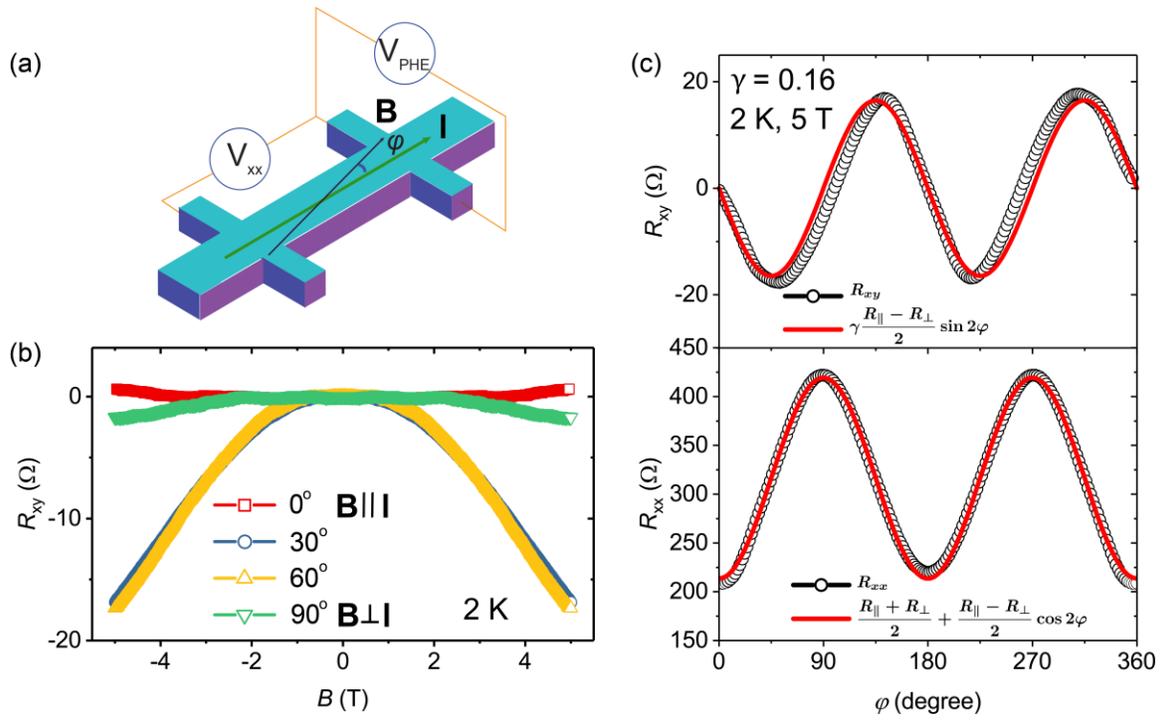

**Figure 2**

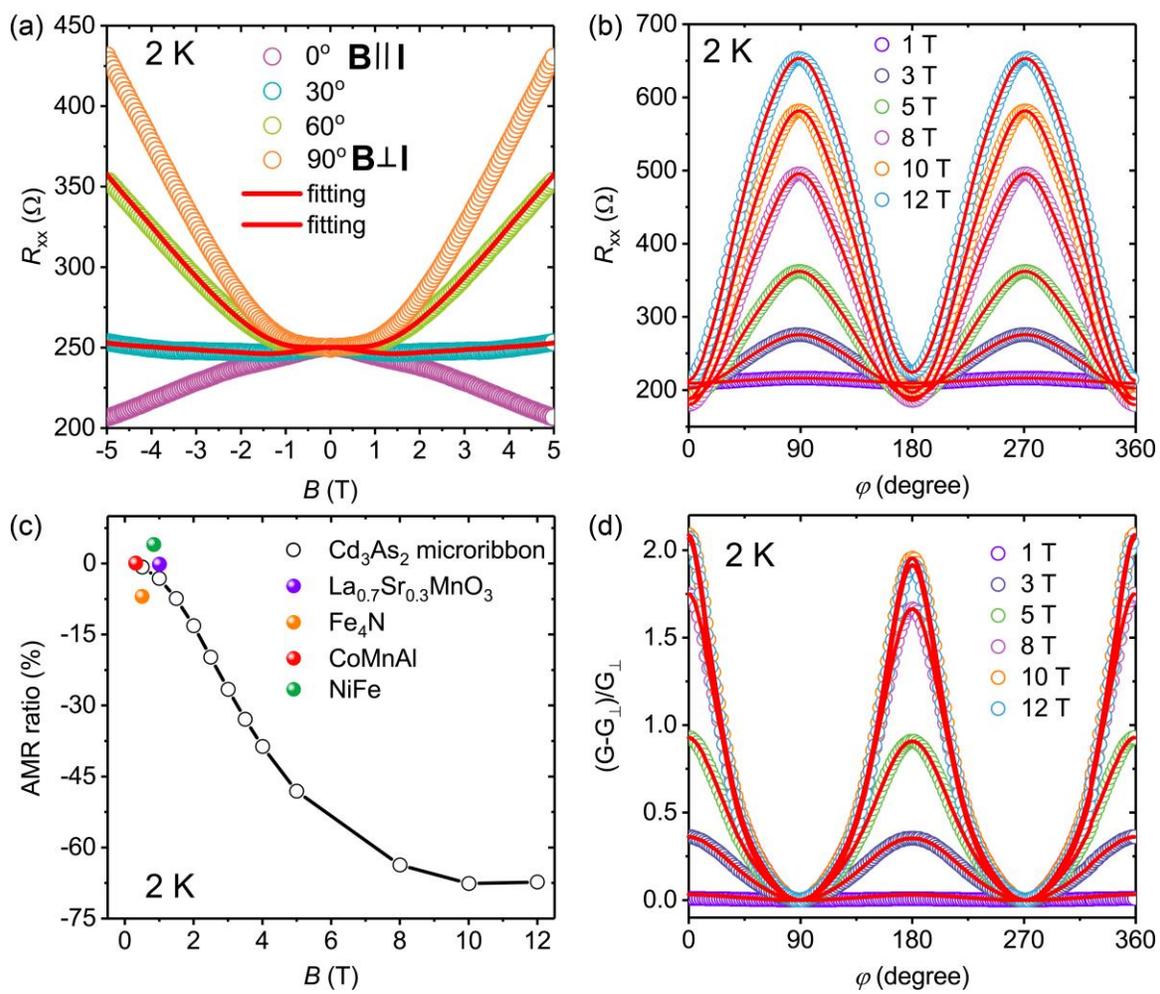

**Figure 3**

# Supplemental Materials

# Giant Anisotropic Magnetoresistance and Planar Hall Effect in the Dirac Semimetal $Cd_3As_2$

Hui Li[1], Huan-Wen Wang[2], Hongtao He[3], Jiannong Wang[1]*, Shun-Qing Shen[2]*

[1]Department of Physics, the Hong Kong University of Science and Technology, Clear Water Bay, Hong Kong, China

[2]Department of Physics, the University of Hong Kong, Pokfulam Road, Hong Kong, China

[3]Department of Physics, Southern University of Science and Technology, Shenzhen, Guangdong 518055, China

## Section A: Magneto-transport characteristics of topological semimetal $Cd_3As_2$ microribbon

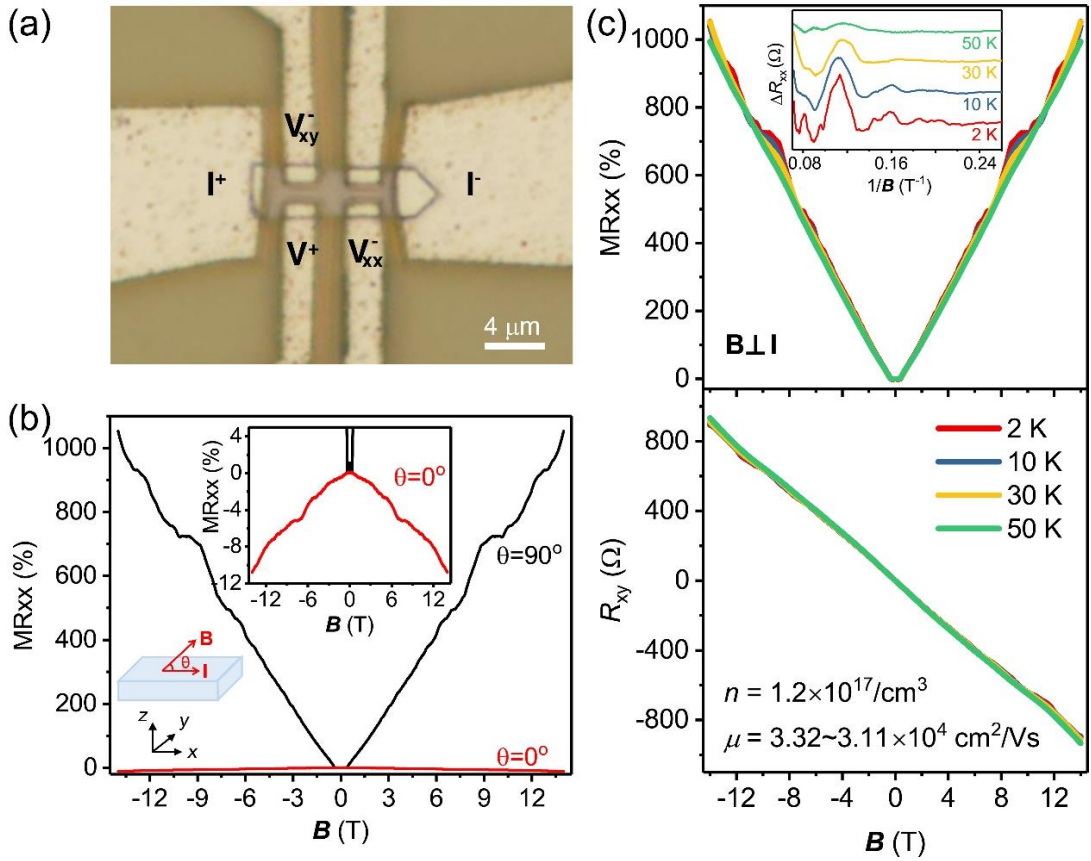

**Figure S1. Magneto-transport characteristics of topological semimetal Cd₃As₂ microribbon.** (**a**) The optical image of a Cd₃As₂ microribbon device. (**b**) The magnetoresistance (MR) measured at 2 K with applied magnetic field (**B**) direction changing from perpendicular ($\theta = 90°$) to parallel ($\theta = 0°$) to the applied current (**I**) direction in *z-x* plane. The inset shows zoom-in view of the negative MR observed when **B**-field parallel ($\theta = 0°$) to the applied current (**I**) direction. (**c**) Magnetoresistance (top panel) and Hall resistance (bottom panel) measured at temperatures indicated with **B**-field direction perpendicular ($\theta = 90°$) to the applied current (**I**) direction in *z-x* plane. The inset in top panel shows the extracted Shubnikov-de-Haas (SdH) oscillations of the Cd₃As₂ microribbon device at temperatures indicated. The calculated carrier concentration and mobility are in the order of $10^{17}$/cm³ and $10^4$ cm²/Vs, respectively.

Figure S1(a) shows the optical image of another typical Cd₃As₂ ribbon device. The width *w* is about 3 μm, the inter-voltage-probe distance for $V_{xx}$ and $V_{xy}$ is about 6 and 1.3 μm, respectively. The magnetoresistance (MR) were measured at *T* = 2 K at different angles ($\theta = 0°$ and 90°) between **B**-field and the applied current (**I**) direction in *z-x* plane (see Figure S1(b)). Similar to the devices measured in the manuscript and our previous work [1], when the **B**-field is applied perpendicular ($\theta = 90°$) to the applied current (**I**) direction, a postive MR up to 1052% at 14 T is observed. On the other hand, a negative MR is obtained when the **B**-field is parallel ($\theta = 0°$) to the applied current (**I**) direction, as it can be seen in the zoom-in view in the inset of Fig. S1(b). In Figure S1(c) the MR curves (top panel) and the conventional Hall resistances (bottom panel) of the Cd₃As₂ ribbon device are shown with **B**-field perpendicular ($\theta = 90°$) to the applied current (**I**) direction at indicated temperatures below 50 K. Shubnikov-de Haas (SdH) oscillations were observed (see inset of top panel in Figure S1(c)), indicating the high quality of our Cd₃As₂ ribbons. A low carrier concentration and a high mobility in the order of $10^{17}$/cm³ and $10^4$ cm²/Vs, respectively, are calculated from the Hall resistancesfor temperatures below 50 K. The Fermi energy $E_F$, defined as the energy difference between the Fermi level and the Dirac point, is estimated to be about 88 meV by using $E_F = \hbar v_F k_F$ where $v_F$ is taken as $1.1 \times 10^6$ m/s according to the ARPES measurements [2].

## Section B: Planar Hall effect (PHE) in Cd$_3$As$_2$ microribbon

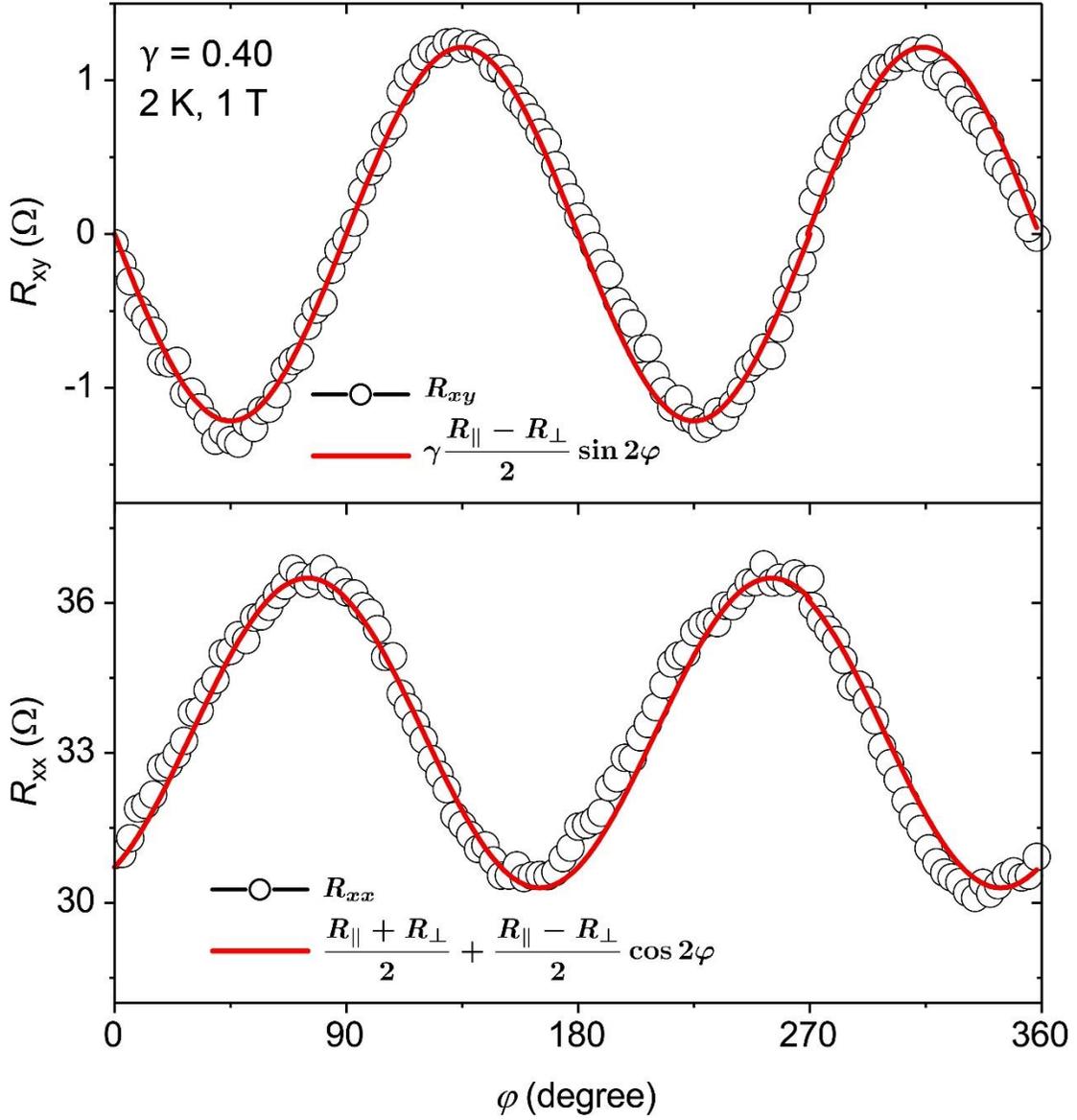

**Figure S2. Planar Hall effect (PHE) and Anisotropic Magnetoresistance (AMR) in Cd$_3$As$_2$ microribbon.** The symmetrized angular dependence of the $R_{xy}$ (top panel) and $R_{xx}$ (bottom panel) are measured at 2 K and 1 T. The red lines are the fitting curves using the inset equations, where $\varphi$ is the angle between the **I** and **B**-field in $x$-$y$ plane, $R_{\parallel}$ and $R_{\perp}$ is the resistance when $\varphi$ equals to 0 and 90°, respectively, $\gamma$ is the ratio of the width to the length of the Cd$_3$As$_2$ microribbon device shown in Figure S1(a).

Figure S2 shows the symmetrized angular-dependent in-plane $R_{xy}$ and $R_{xx}$ measured at 2 K and 1 T of the Cd$_3$As$_2$ microribbon device shown in Figure S1(a). Similar to the device measured in the main text, both the measured $R_{xy}$ and $R_{xx}$ show an 180° periodic angular dependence, which is in agreement with Equation (1) in the main text. The $R_{xy}$ was fitted using the equation $R_{xy} = \gamma \frac{R_\parallel - R_\perp}{2} \sin 2\varphi$ derived from Equation (1) in the main text, where $R_\parallel$ and $R_\perp$ is the in-plane longitudinal resistance when $\varphi$ equals to 0 and 90°, respectively, and $\gamma$ is the geometric ratio of the width to the length of the Hall bar device. Remarkably, as indicated by the red line in the top panel in Figure S2, the measured in-plane $R_{xy}$ can be well fitted by the equation, demonstrating existence of the PHE in this Cd$_3$As$_2$ microribbon device. Moreover, the measured in-plane $R_{xx}$ can also be well fitted by the equation $R_{xx}(B,\varphi) = \frac{R_\parallel + R_\perp}{2} + \frac{R_\parallel - R_\perp}{2} \cos 2\varphi$, as indicated by the red line in the bottom panel in Figure S2, which is a peculiar characteristic of the AMR as discussed in the main text.

## Section C: B-field dependence of AMR and PHE

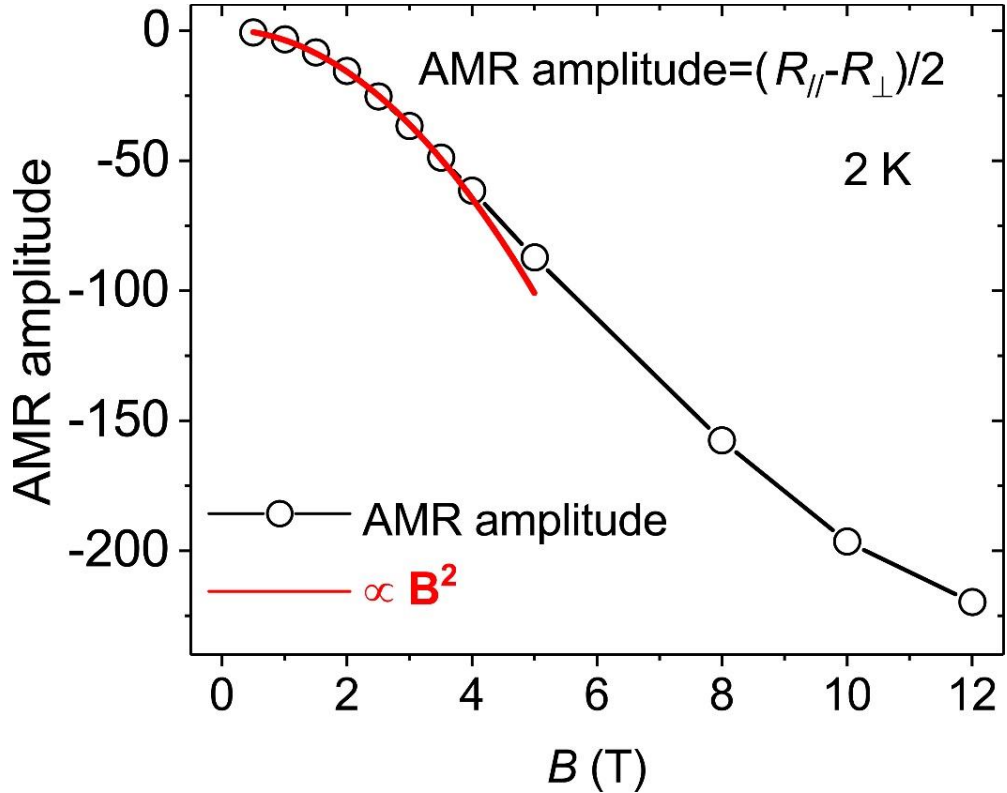

**Figure S3. B-field dependence of AMR amplitude at 2 K.** The AMR amplitude, defined as $\frac{R_\parallel - R_\perp}{2}$, as a function of **B**-fields at 2 K is shown. The open circles are the experimentally measured data. The red curve shows the AMR amplitude can be well fitted by a quadratic function at small **B**-field regime ($B < 1.0$ T). The black line is a guide for the eyes.

The AMR and PHE follow the same **B**-field dependence. As shown in Figure S3, we have plotted the **B**-field dependence of AMR amplitude, defined as $\frac{R_\parallel - R_\perp}{2}$. With increasing **B**-fields, the AMR amplitude increases monotonically and reaches to about -225 at 12 T and 2 K. Theoretically, due to the constraint of Onsager relation, the leading order magnetoresistance (MR) should be of $B^2$ for a system with time reversal symmetry. Thus, the AMR amplitude is expected to follow the quadratic **B**-field dependence at small **B**-field regime.

As indicated by red curve in Figure S3, the AMR amplitude can be well fitted by a quadratic function at small **B**-field regime ($B < 1.0$ T), which is the same as the calculations from the semiclassical transport theory [3,4]. However, when the applied **B**-field increases over 3.5 T, the AMR amplitude deviates from the quadratic function, which may be caused by multiple reasons and is still under theoretical investigation.

## Section D: Theoretical calculations of AMR and PHE

The longitudinal resistance $R_{xx}$ and Hall resistance $R_{xy}$ of the devices can be calculated by

$$R_{xx} = \frac{V_{xx}}{I_x} = \frac{E_x l}{j_x w t} = \rho_{xx} \frac{l}{wt} \tag{1}$$

$$R_{xy} = \frac{V_{xy}}{I_x} = \frac{E_y w}{j_x w t} = \frac{E_y}{j_x t} = \frac{\rho_{xy}}{t} \tag{2}$$

where the $l$, $w$ and $t$ are the length, width and thickness of the devices, respectively.

According to AMR and PHE theory, the in-plane field dependent resistivity is given by $\rho_{ij} = \rho_\perp \delta_{ij} + (\rho_\parallel - \rho_\perp) B_i B_j / B^2$ with $i, j = x, y$, i.e. $\rho_{xx} = \frac{\rho_\parallel + \rho_\perp}{2} + \frac{\rho_\parallel - \rho_\perp}{2} \cos 2\varphi$, and $\rho_{xy} = \frac{\rho_\parallel - \rho_\perp}{2} \sin 2\varphi$, where $\varphi$ is the angle between the magnetic field and electric current density.

By replacing $\rho_{xx}$ and $\rho_{xy}$ in the equation (1) and (2), we have

$$R_{xx} = \frac{l}{wt}\left(\frac{\rho_\parallel + \rho_\perp}{2} + \frac{\rho_\parallel - \rho_\perp}{2}\cos 2\varphi\right) = \frac{R_\parallel + R_\perp}{2} + \frac{R_\parallel - R_\perp}{2}\cos 2\varphi \quad (3)$$

$$R_{xy} = \frac{\rho_\parallel - \rho_\perp}{2t}\sin 2\varphi = \frac{wt}{l}\frac{R_\parallel - R_\perp}{2t}\sin 2\varphi = \gamma\frac{R_\parallel - R_\perp}{2}\sin 2\varphi \quad (4)$$

where $\gamma$ is the geometric ratio of the width to the length of the Hall bar device.